\documentclass[twocolumn,showpacs,prl,amsmath,amssymb,floatfix,superscriptaddress]{revtex4} 
\usepackage{graphicx}% Include figure files
\usepackage{dcolumn}% Align table columns on decimal point
\usepackage{bm}% bold math

\begin{document} 
% \draft command makes pacs numbers print 

\setlength{\topmargin}{0in}

\title{Evidence of decoupled lattice distortion and 
ferroelectric polarization in the relaxor system PMN-$x$PT} 
\author{Guangyong Xu}
\affiliation{Physics Department, Brookhaven National Laboratory, Upton, 
New York 11973}
\author{D.~Viehland}
\affiliation{Department of Materials Science and Engineering, Virginia 
Tech., Blacksburg, VA 24061}
\author{J.~F.~Li}
\affiliation{Department of Materials Science and Engineering, Virginia 
Tech., Blacksburg, VA 24061}
\author{P.~M.~Gehring}
\affiliation{NIST Center for Neutron Research, National Institute of Standards
and Technology, Gaithersburg, Maryland, 20899}
\author{G.~Shirane} 
\affiliation{Physics Department, Brookhaven National Laboratory, Upton, 
New York 11973}
\date{\today} 
 
\begin{abstract} 

We report high $q$-resolution neutron scattering data on PMN-$x$PT single 
crystals with $x=20\%$ and $27\%$. No rhombohedral distortion occurs
in the 20PT sample for temperatures as low as 50~K.  On the other hand, 
the 27PT sample transforms into a rhombohedral phase below 
$T_C \sim 375$~K. Our data provide conclusive evidence that a new phase with an 
average cubic lattice is present in the bulk of this 
relaxor system at low PT concentration, in which the ferroelectric polarization 
and lattice distortion are decoupled. The rhombohedral distortion is limited to
the outermost tens of microns of the crystal.

\end{abstract} 
 
\pacs{77.80.-e, 77.84.Dy, 61.12.Ex} 

\maketitle

%\section{Introduction}
% 
%
%\section{Experiment}
%
%\section{Discussion}

\begin{figure}[ht]
\includegraphics[width=\linewidth]{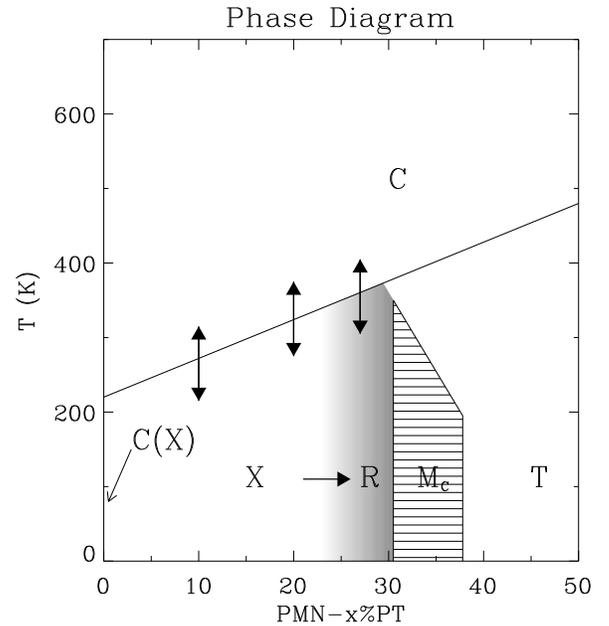}
\caption{Schematic of the revised phase diagram of PMN-$x$PT in zero field.
The arrows indicate compositions that have been studied using high 
$q$-resolution neutron scattering techniques.}
\label{fig:fig1}
\end{figure}

The complex perovskite system  Pb(Mg$_{1/3}$Nb$_{2/3}$)O$_3$ (PMN) is a 
close analogue to Pb(Zn$_{1/3}$Nb$_{2/3}$)O$_3$ (PZN), both of which have been 
studied extensively because of their extraordinary piezoelectric 
properties. The addition of PbTiO$_3$ (PT) enhances the piezoelectricity 
significantly, forming solid solutions PMN-$x$PT and PZN-$x$PT for 
$0 \le x \le 1$~\cite{PZT1,PZN_phase2}.  This 
discovery has motivated numerous studies of the structural phase 
transitions in these systems as a function of $x$. The current zero-field 
phase diagrams of PMN-$x$PT and PZN-$x$PT share many common features, 
including a purported cubic-to-rhombohedral phase transition at small 
values of $x$~\cite{PZN_phase2,PMN_phase,PZN_phase} (see Fig.~\ref{fig:fig1}).  
One exception to this is pure PMN ($x=0$), 
which is reported to retain an average cubic structure down to 
temperatures as low as 5~K~\cite{Bonneau,Husson}.  
The rhombohedral distortions in both 
PMN-$x$PT and PZN-$x$PT have been observed directly by x-ray  
scattering experiments~\cite{PMN_phase,PMN_Ye, Lebon} and related to the 
establishment of ferroelectric order in each system.

Several recent studies have now uncovered evidence of a new phase located 
on the rhombohedral side of the phase diagram. 
Electric field studies of PZN-8PT by 
Ohwada~{\it et al.}~\cite{PZN_efield} show that the unpoled system 
does not exhibit a rhombohedral unit cell distortion below $T_C$. 
This was the first hint that 
a new phase, which was labeled phase X, may exist there. 
Subsequent high energy x-ray structural studies by 
Xu {\it et al.}~\cite{PZN_Xu} prove that the inside of single crystal PZN has 
an undistorted lattice, providing conclusive evidence of a new phase. 
Motivated by these findings, Gehring {\it et al.}~\cite{PMN-10PT} performed
neutron scattering measurements with very fine q-resolution on a single crystal 
sample of PMN-10PT. In contrast to previous 8.9~keV Cu K$_\alpha$ x-ray 
measurements by Dkhil {\it et al.}~\cite{PMN_diffuse2} on 
a powder sample of the same compound, 
the neutron data show no evidence of a rhombohedral distortion below $T_C$, 
and are consistent with the presence of phase X. 

In this Letter we present high q-resolution neutron scattering data on
PMN-$x$PT with $x=20\%$ (20PT) and $27\%$ (27PT),
both of which are on the the R(X) side of the phase diagram 
(See Fig.~\ref{fig:fig1}). Our data show that these two compounds are
notably different. One of them (20PT) exists as phase X at low temperatures, 
which has an undistorted lattice, inconsistent with the previously reported 
rhombohedral distortion by x-ray measurements on similar 
compounds~\cite{PMN_Ye}. On the other hand, 27PT shows a clear rhombohedral 
distortion below $T_C \sim 375$~K. We believe that phase X is one  in which a
rhombohedral polarization exists within an undistorted lattice,
and thus highly unusual. This decoupling breaks down with
increasing the PT concentration $x$, and allows
the local atomic displacements to develop into a global rhombohedral phase.

Our samples are high quality single crystals of PMN-20PT and 27PT, 
having dimensions
$3\times3\times2$~mm$^3$ and $7\times3\times2$~mm$^3$, respectively.
Both of them have been depoled before the neutron measurements.
The neutron scattering experiments were performed on the BT9 triple-axis
spectrometer at the NIST Center for Neutron Research. Measurements were made 
using a fixed incident neutron energy
$E_i$ of 14.7~meV, obtained from the (002) reflection of a PG
monochromator, and horizontal beam collimations of 10'-46'-20'-40'. 
We exploited the (004) reflection of a perfect Ge crystal as
analyzer to achieve unusually fine q-resolution near the relaxor (220)
Bragg peak thanks to a nearly perfect matching of the sample and analyzer
d-spacings. Close to the (220) Bragg peak, 
the $q$-resolution along the wave vector direction is about  
0.0012~\AA$^{-1}$ ($\delta q/q \approx 2\times 10^{-4}$)~\cite{High_q}. 

\begin{figure}[ht]
\includegraphics[width=\linewidth]{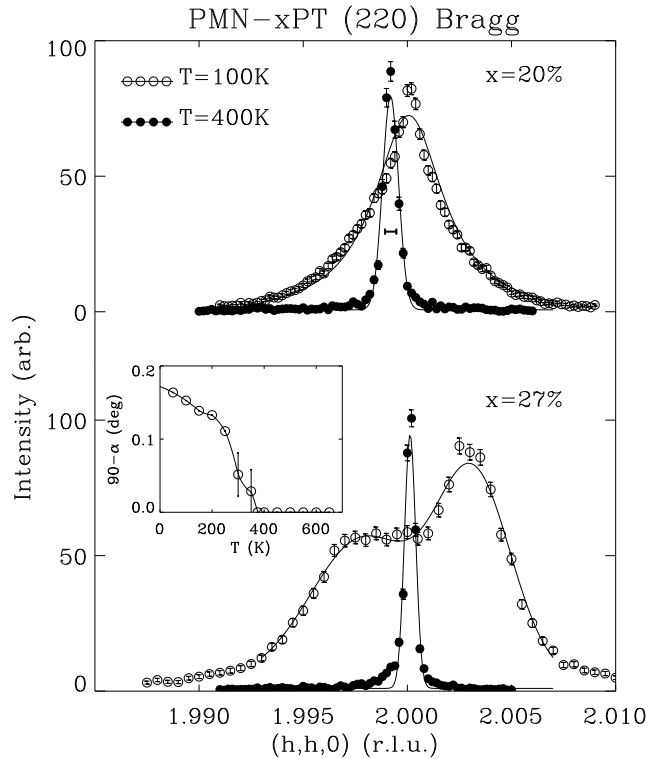}
\caption{Profiles of the (220) Bragg peaks for 20PT (top panel) and 27PT
(bottom panel) are shown at 100 K (open circles) and 500 K (solid
circles).  The solid lines are fits described in the text. The horizontal bar 
indicates the $q$ resolution along the (220) direction.
The inset shows the temperature dependence of the rhombohedral
angle $\alpha$ for the 27PT sample.}
\label{fig:fig2}
\end{figure}

Fig.~\ref{fig:fig2} shows longitudinal ($\theta-2\theta$) scans along 
the (220) pseudocubic Bragg peak both above and below $T_C$. 
The relative intensities have been scaled for ease of comparison. 
The Bragg profile for 20PT can be fit to a single peak at all
temperatures and shows no evidence of any rhombohedral splitting.  Above
$T_C \sim 300$~K the profile fits well to a Gaussian line-shape with a narrow
width only slightly larger than that of the instrumental resolution.
However the Bragg peak broadens considerably below $T_C$ and is better
described by a Lorentzian function.  This broadening indicates the
presence of a large internal strain within the crystal bulk below $T_C$.

By contrast, 27PT has a rhombohedral structure at low temperature. 
The (220) Bragg peak splits below $T_C \sim 375$~K,
and is an unambiguous sign of the expected rhombohedral phase transition.
The rhombohedral distortion increases on cooling, as shown by the  
plot of the rhombohedral angle $\alpha$ vs T (inset of Fig.~\ref{fig:fig2}). 

\begin{figure}[ht]
\includegraphics[width=\linewidth]{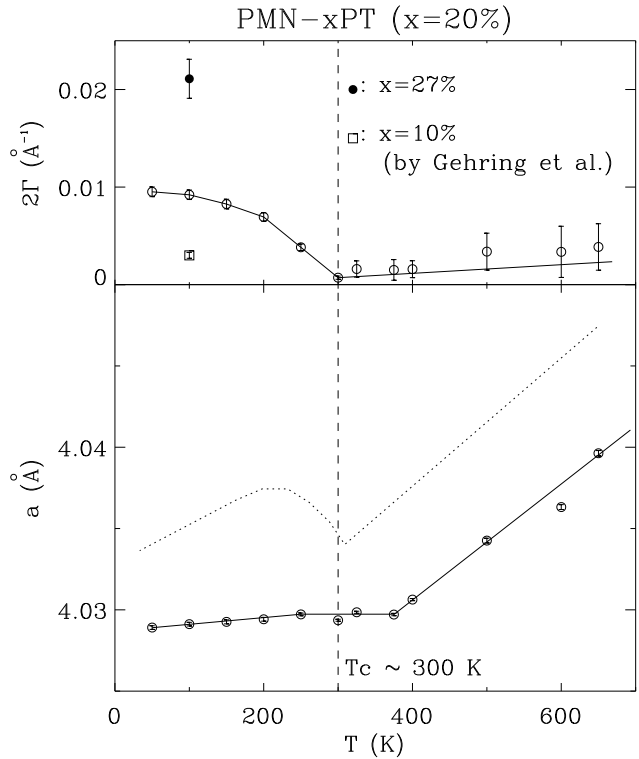}
\caption{Top panel: the resolution corrected FWHM (2$\Gamma$) of the (220) 
Bragg peak vs T for 20PT (open circles), compared with data from 27PT 
(close circle) and 10PT (square) by Gehring {\it et al}~\cite{PMN-10PT}. 
Bottom panel: lattice parameter $a$ vs T. The dotted line represents thermal 
expansion behavior typical of normal ferroelectrics. The solid lines are guides to the eye.}
\label{fig:fig3}
\end{figure}

These results clearly indicate that the low temperature phase of bulk 20PT 
is different from that of the rhombohedral phase observed in
27PT.  The longitudinal full width at half maximum ($2\Gamma$) of the (220)
Bragg peak is plotted versus temperature in Fig.~\ref{fig:fig3} as is the 
lattice parameter $a$.  Data taken on heating and cooling yield similar 
results and thus no hysteretic effects.
The 20PT Bragg profile is sharp and almost resolution limited at
high temperature, but begins to broaden below $T_C \sim 300$~K, which is 
indicative of a true phase transition. 
For comparison, we also show the values of $2\Gamma$ at 100 K for
the 27PT sample (solid circle) and  the PMN-10PT sample (open rectangle)
measured by Gehring {\it et al.}~\cite{PMN-10PT}. 
Clearly the broadening develops not only with cooling, 
but also with increasing PT concentration.

For normal ferroelectric oxides, the lattice parameter decreases linearly on 
cooling down to $T_C$, then increases in the ferroelectric 
region, as shown by the dotted line in Fig.~\ref{fig:fig3}.
In relaxors, the linearity breaks down at a temperature $T_x$ above $T_C$.
Experimental evidence for this have been provided  by the 
independent measurements
of Dkhil {\it et al.}~\cite{PMN_diffuse2} and Ye {\it et al.}~\cite{PMN_Ye} on 
a series of PMN-$x$PT samples. 
The 20PT lattice parameter follows the linear behavior seen in
conventional ferroelectrics at high temperature, but shows a break at 
$T_x \sim 380$~K, and remains almost constant below $T_x$.  
In addition, the unit cell
volume does not increase with cooling in the ferroelectric region.  This
anomalous feature is characteristic of phase X, and has also been observed
in pure PMN~\cite{PMN_Zhao, PMN_diffuse2} and PMN-10PT~\cite{PMN-10PT}.

Our results provide additional evidence that PMN-xPT transforms into
phase X at low temperature for small PT concentrations.  The studies by
Ohwada {\it et al.}~\cite{PZN_efield} and Xu {\it et al.}~\cite{PZN_Xu} 
provide compelling evidence that the same picture is true for PZN-xPT.  
By using x-rays with different incident energies, 
Xu {\it et al.}~\cite{PZN_Xu}  also found out that the outermost 
$\sim50$~$\mu$m of the PZN crystal does exhibit a rhombohedral distortion.  
The thickness of the outer-layer is estimated from the penetration depths
of different x-ray energies, and may vary with different compounds. 
However, it is important to note that the existence of an outer-layer with 
a structure different from that of the inside naturally explains the 
discrepancies between earlier x-ray powder diffraction and the recent neutron 
and high-energy x-ray scattering results. And the fact that x-ray studies 
of pure PMN have not observed a rhombohedral phase could be understood if PMN
represents the limiting case of phase X with a very thin outer-layer.

Our results, together with other neutron scattering measurements of
phase X~\cite{Stock1,PMN-10PT}, show that all measured Bragg peaks broaden 
significantly in the longitudinal direction below the Curie temperature $T_C$. 
These neutron results are in contrast to the high energy x-ray
case~\cite{PZN_Xu}, where the Bragg profiles are sharp for temperatures both 
above and below $T_C$.
We speculate that  these differences are due to the fact the incident 
neutron beam has a much larger size and divergence than the x-ray beam,
therefore more sensitive to the inhomogeneity of the whole crystal.
More measurements are currently underway for better understanding of these 
differences.

In phase X, the crystal lattice are undistorted, and  the Bragg peaks do not 
split in diffraction measurements. However, the true symmetry is not yet known.
To get detailed information on the atomic positions, 
we are planning further investigations on refinements from single crystal 
diffractions. Unfortunately, extinction effects arising from those near-perfect 
crystals are making this quite difficult.

\begin{figure}[ht]
\includegraphics[width=\linewidth]{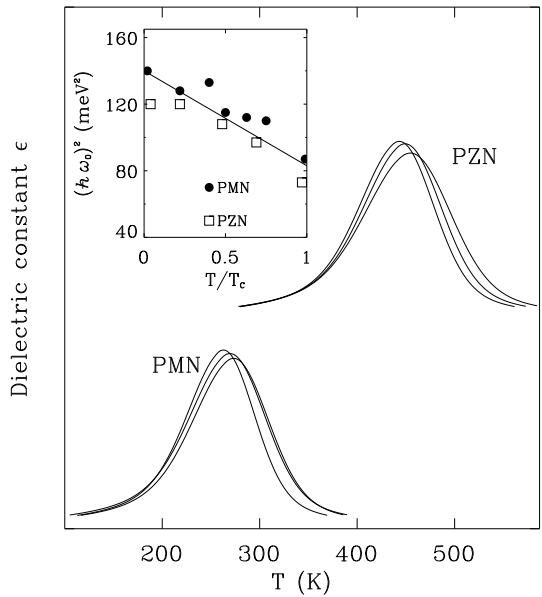}
\caption{A schematic comparison between PMN and PZN, showing the frequency 
dependent dielectric constant $\epsilon$ vs T. The inset shows the soft TO 
phonon energy squared vs T/$T_C$ for PMN~\cite{Waki1} and 
PZN~\cite{Stock1}.}
\label{fig:fig4}
\end{figure}

Both PMN and PZN are relaxors that exhibit a broad and strongly 
frequency-dependent dielectric constant $\epsilon$ as shown in 
Fig.~\ref{fig:fig4}. Despite the absence of a rhombohedral distortion in
phase X, ferroelectric polar order is still present in the system. 
Recently, neutron inelastic scattering measurements by 
Wakimoto {\it et al.}~\cite{Waki1} of the ferroelectric soft TO phonon 
in pure PMN revealed that the soft mode recovers at 
temperatures below $T_C \sim 213$~K. 
Similar measurements on PZN were later performed by
Stock {\it et al.}~\cite{Stock1}, where $T_C \sim 410$~K.  
The linear relationship between the soft TO phonon 
energy squared $(\hbar\omega_0)^2 \propto 1/\epsilon$ and T 
(see inset of Fig.~\ref{fig:fig4}) is a clear signature of 
an ordered ferroelectric phase below $T_C$. 

    In ferroelectric systems, the primary order parameter is the
polarization $\vec{P}$ caused by ionic shifts. It is usually coupled to 
a secondary order parameter - the lattice distortion, which changes the shape 
of the unit cell and is more readily  accessible by scattering methods. 
Recent reports on epitaxial films of SrTiO$_3$ and 
BaTiO$_3$~\cite{SrTiO3} show that this coupling is broken in these 
systems, possibly due to epitaxy strain and substrate clamping effects.
The coexistence of ferroelectric spontaneous polarization 
and an undistorted  lattice is the first example of the decoupling inside a 
free-standing crystal.

In pure PMN, polar regions with sizes of a few nanometers
start to form at the Burns temperature~\cite{Burns} $T_d \approx 600$~K, 
and contribute directly to diffuse scattering~\cite{PMN_neutron,PMN_neutron2}. 
Both neutron and x-ray 
measurements~\cite{PMN_diffuse2,PMN_xraydiffuse,PZN_diffuse,Koo}
show that in relaxors, the diffuse scattering intensity starts to be visible 
around $T_d$, and increases monotonically on cooling.
An important step in understanding the nature of these polar nanoregions (PNR) 
has been made by Hirota {\it et al.}~\cite{PMN_diffuse}, with the concept of 
phase-shifted PNR. The atomic displacements derived from previous neutron 
diffuse scattering data on PMN~\cite{PMN_neutron3} by Vakhrushev {\it et al.} 
were re-examined and decomposed into
the sum of two terms with comparable magnitudes. The first term satisfies the
center-of-mass condition and is consistent with the values derived from
inelastic scattering intensities from the soft TO 
mode~\cite{Waki1,PMN_softmode}. The second term 
represents a uniform  shift along the polar directions of the PNR relative to 
the surrounding lattices. In this picture, the PNR condense from the soft TO 
mode, and exhibit a "uniform phase shift", or displacement, below $T_d$.
Experimental evidence of this uniform phase shift has been observed by 
Gehring {\it et al.}~\cite{PZN-8PT} by applying an external electric field 
along the [001] direction of a PZN-8PT single crystal. The diffuse scattering 
intensity around the (003) Bragg peak is considerably reduced at all 
temperatures below $T_C$, while the diffuse scattering around (300) remains 
strong. The size of this phase shift is comparable to the first term and  
is estimated to be around 1/10 of the lattice spacing. This large phase 
shift creates a huge energy barrier, making it extremely difficult for the PNR 
to merge into the surrounding lattice.

The interaction between the PNR and the surrounding lattice can be viewed as a 
coupling parameter between the ferroelectric polarization and lattice 
distortion. We conjecture that the balance between the energy barrier  
caused by the uniform phase shift and this coupling parameter 
is the key in understanding phase X.
For $T < T_C$, the entire system is polarized.
Nevertheless, in order for a long range rhombohedral distortion to develop,
the energy barrier must be overcome before the PNR can 
merge and grow into macroscopic rhombohedrally distorted domains. 
This can only be achieved
when the interaction between the PNR and the surrounding environment is 
sufficiently 
strong. Otherwise, the lattice will remain undistorted, with confined
ferroelectric polarization. In other words, 
phase X is a special confined form of rhombohedral phase. Electric field
studies on PZN-8PT by Ohwada {\it et al}~\cite{PZN_efield} confirm this 
by showing that phase X always transforms into the M$_A$ phase first with 
increasing field, following the polarization rotation path of 
$R \rightarrow M_A \rightarrow M_C \rightarrow T$, indicating a \{111\} 
rhombohedral type of polarization in phase X. 

The confinement of local rhombohedral order in phase X inevitably induces 
instability in the system. The interaction between the PNR and the surrounding 
undistorted lattice creates strain and stress in the system, which leads to the 
broadening of the Bragg profile below $T_C$.
A small external ``force'' may be able to drive the whole system into a more 
stable phase. For example, by application of a small electric field, even pure 
PMN can be transformed into a rhombohedral phase below $T_C$~\cite{Calvarin}. 
With increasing PT concentration, the coupling becomes stronger and the 
instability increases. Eventually, the ``confinement'' breaks down and the 
usual coupling between rhombohedral distortion and 
ferroelectric polarization is re-established for larger PT concentrations.

\begin{acknowledgments}
We would like to thank C.~Stock and S.~B.~Vakhrushev
for stimulating discussions, and H.~C. Materials for providing the 
single crystals. Financial support from the U.S. Department of 
Energy under contract No.~DE-AC02-98CH10886 and Office of Naval Research 
under contracts N000140210340, N000140210126, and MURI N000140110761
is also gratefully acknowledged. 
\end{acknowledgments}

%\section{Acknowledgments}
%\bibliography{PZN}

\begin{thebibliography}{28}
\bibitem[{Hig()}]{High_q}
\bibinfo{note}{Details of the high $q$-resolution neutron scattering technique
  will be discussed elsewhere.}
\end{thebibliography}

\end{document}